\documentclass[
 aip,
 long, numerical bibliography, (default)
 jmp,%
 amsmath,amssymb,
preprint,%
nofootinbib]{revtex4-1}

\usepackage[a4paper,margin=2.5cm]{geometry}
\usepackage[utf8]{inputenc}
\usepackage{amsmath}
\usepackage{graphicx}% Include Figure files
\usepackage{dcolumn}% Align table columns on decimal point
\usepackage{bm}% bold math
\usepackage{epstopdf}
\usepackage{hyperref}
\usepackage{pdfpages}

\makeatletter
\AtBeginDocument{\let\LS@rot\@undefined}
\makeatother

\begin{document}

\title{Tension mediated nonlinear coupling between orthogonal mechanical modes of nanowire resonators}

\author{John P. Mathew}
\affiliation{Department of Condensed Matter Physics and Materials Science, Tata Institute of Fundamental Research, Homi Bhabha Road, Mumbai 400005 India}
\author{Anand Bhushan}
\email{anand.bhushan@nitp.ac.in}
\affiliation{Department of Mechanical Engineering, National Institute of Technology, Patna 800005 India}
\author{Mandar M. Deshmukh}
\email{deshmukh@tifr.res.in}
\affiliation{Department of Condensed Matter Physics and Materials Science, Tata Institute of Fundamental Research, Homi Bhabha Road, Mumbai 400005 India}

\begin{abstract}
We study the nonlinear coupling between orthogonal flexural modes of doubly clamped InAs nanowire resonators. The two orthogonal modes are formed by the symmetry breaking and lifting of degeneracy of the fundamental mode. The presence of a Duffing nonlinearity emerges when a mode is driven to large amplitudes. In this regime the modes are coupled due to the tension induced from the large amplitude of oscillations and is reflected in the hysteretic response of the mode that is not strongly driven. We study the driven-driven response of the mechanical modes to elucidate the role of nonlinear mode coupling in such mechanical resonators. The dynamics of the coupled modes studied here could prove useful in technological applications such as nanowire based vectorial force sensing.
\end{abstract}

\maketitle

\section{Introduction}
Understanding the coupling between different modes of a complex system is of great interest as it can be used to improve functionality such as enhancing the sensitivity of the system. Optomechanics, for example, has harnessed the coupling between a mechanical oscillator and an optical cavity \cite{aspelmeyer_cavity_2014} to explore the limits of precise position measurements. %The underlying idea of coupling between a mechanical oscillator and a cavity can be generally extended to completely mechanical systems with coupling between two vibrational modes of the system where the second mode serves as the cavity \cite{mahboob_phonon-cavity_2012,okamoto_coherent_2013}.
Similarly, the coupling between two vibrational modes of nanomechanical systems has also been explored in recent times.

Coupling between mechanical modes has been studied in nanofabricated doubly clamped beams \cite{mahboob_phonon-cavity_2012,okamoto_coherent_2013,faust_nonadiabatic_2012},  nanofabricated cantilevers \cite{westra_interactions_2011,westra_nonlinear_2010}, carbon nanotubes \cite{castellanos-gomez_strong_2012, eichler2012strong}, 2D materials \cite{alba_tunable_2016,mathew_dynamical_2016,liu_optical_2015}, cantilevers of VLS (vapour liquid solid) grown nanowires  \cite{cadeddu_time-resolved_2016,lepinay_universal_2016,foster_tuning_2016,rossi_vectorial_2016} and doubly clamped VLS nanowires \cite{solanki_tuning_2010}. The origin of nonlinear intermodal coupling can be complex and system specific \cite{lifshitz2008nonlinear,khan_tension-induced_2013,eriksson_frequency_2013}; efforts to understand the microscopic origin are ongoing across systems as tunability of this nonlinear coupling can provide additional functionality to devices.

VLS grown nanowires \cite{dasgupta_25th_2014} offer advantage of engineering the properties with unprecedented control \cite{algra_twinning_2008,johansson_controlled_2009}.
%Bottom-up synthesis of nanowires has been used to realize coupling between quantum dot structure and mechanics \cite{montinaro_quantum_2014,auffeves_strain-mediated_2014}.
Recently, orthogonal vibrational modes of the cantilevers, singly clamped nanowires, made using VLS grown nanowires have been used to vectorially map \cite{lepinay_universal_2016,rossi_vectorial_2016} the forces on a surface to provide maps of electric fields. The two orthogonal vibrational modes arise due to an unintentional breaking of symmetry of the degenerate fundamental mode. This pair of modes can provide complementary information about the nano electromechanical system (NEMS) and are, hence, of interest. We show that the Duffing nonlinearity of a mode that is driven to large amplitudes imprints itself as a hysteretic increase in frequency of the second mode. In this paper, we study in detail the inter-modal nonlinear coupling and show that it can be large and tunable as a function of the tension in the nanowires. Tunability due to tension is absent in singly clamped beams where nonlinear coupling has been studied recently \cite{cadeddu_time-resolved_2016}. We compare the strength of coupling across systems and find that the coupling in our system is large and comparable to the coupling observed in carbon nanotube resonators. In addition, the key role of tension mediated intermodal coupling is demonstrated by the fact that the strength of the coupling in our doubly clamped nanowire devices is significantly larger than cantilever devices made using nanowires. The microscopic model we develop explains the experimental observations accurately.

\section{Experimental methods}
We start by briefly describing the nanofabrication of devices used in our experiments; details of fabrication are available in previous reports \cite{solanki_tuning_2010,abhilash2012}. InAs nanowires of $\sim$10 $\mu$m length and 100 nm diameter were grown using MOCVD techniques. Intrinsic silicon wafers with 100 nm of thermally grown nitride were used as the substrate for device fabrication. A thin layer of electron beam (e-beam) resist was spun on the substrate. The thickness of this resist layer decides the height above the substrate by which the nanowire is suspended.
The nanowires were removed from the growth substrate and then drop cast above the e-beam resist coated substrate and then covered by additional layers of e-beam resists.
After patterning source, drain, and gate electrodes using standard e-beam lithography, the substrate was loaded in to a sputtering system with an \textit{in-situ} plasma etcher for conformal deposition of metal. Prior to metal deposition the sample was exposed to an argon plasma to remove the residual oxide on the nanowires. This allows the formation of Ohmic contact to the nanowire. Gold was sputtered on the substrate preceded by a thin layer of chromium for adhesion. The sputtered metal forms both the electrical contacts as well as mechanical anchors for suspending the device. Lift-off followed by drying the substrate completes the device fabrication process.

\begin{figure}
\centering
\includegraphics[width=7.5cm]{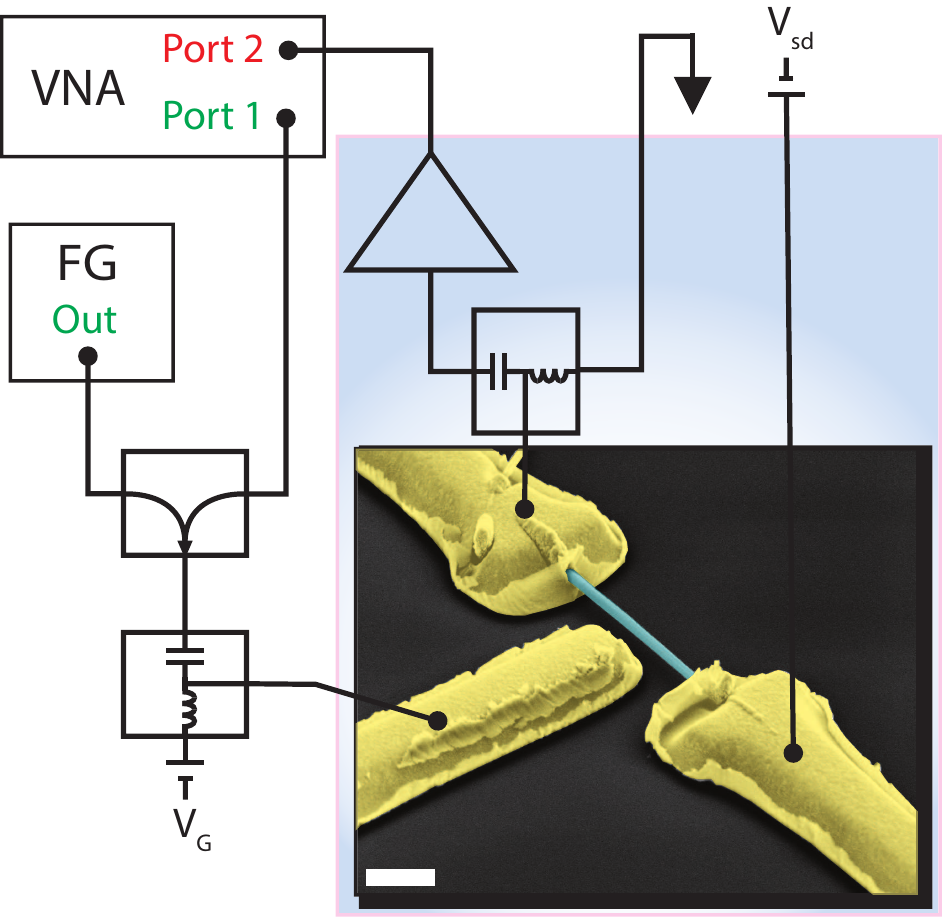}
\caption{\textbf{Circuit schematic.} \label{circuit}SEM image of the device along with a schematic of the circuit used in our experiments is shown here. $V_{sd}$ is applied to the source electrode and the \textit{rf} current through the nanowire is amplified and measured through the drain electrode. The gate electrode is positioned $\sim$ 500 nm away from the nanowire. VNA: vector network analyzer, FG: function generator. The blue shaded area denotes the parts inside the cryostat. Scale bar corresponds to 2 $\mu$m.}
\end{figure}

Figure \ref{circuit} shows a false colored scanning electron microscope (SEM) image of a device along with the schematic of the circuit used in our experiments.
The insulating nature of the intrinsic silicon substrates at low temperatures negates the effects of parasitic capacitances and allows us to carry out direct readout of the electrical signal at radio frequencies (\textit{rf})\cite{abhilash2012,xu2010radio}. DC and \textit{rf} signals were combined using a bias tee and applied to the gate electrode to actuate the mechanical oscillations in the resonator.  The change in \textit{rf} current through the device due to mechanical oscillations were then measured by a vector network analyzer (VNA). The signal from the VNA was used to drive and probe one mechanical mode in the linear regime, while another signal combined from a function generator (FG) was used to drive and pump the second mode to large amplitude oscillations. This allows us to investigate the driven response of one mechanical mode to large amplitude oscillations of the second mode. All experiments were performed with sample temperature at $\sim$5~K.

\begin{figure}
\centering
\includegraphics[width=7.5cm]{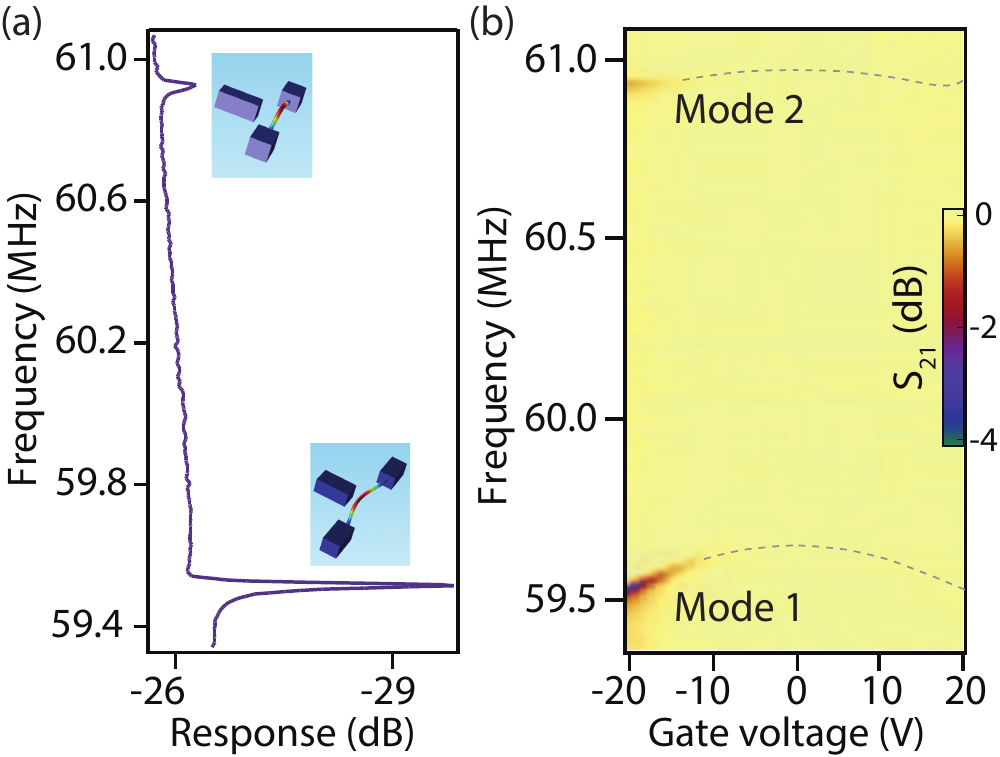}
\caption{\textbf{Electromechanical response.} \label{gating} (a) The two modes of the resonator are separated by $\sim$ 1.4 MHz for an applied DC gate voltage of -20 V. (b)  Dispersion of the two modes with DC gate voltage. Color plot shows the calibrated response obtained from the VNA. Dashed lines are guides to the eye. Power used to drive the resonators is -30 dBm in both (a) and (b).}
\end{figure}

Figure \ref{gating}(a) shows the response of the nanowire resonator measured using the network analyzer with the pump signals turned off for an applied DC gate voltage of -20 V. The lower frequency mode at $\sim$ 59.5 MHz is seen to have a large signal compared to the higher mode at $\sim$ 60.9 MHz with quality factor of $\sim$2800 for both modes. Here, the amplitude of mode 1 at resonance is estimated to be $\sim$ 0.8 nm (see supplementary material) for a drive power of -30 dBm. For the remainder of the article we shall refer to the lower/higher frequency mode as mode 1/mode 2. The small frequency difference ($\sim$1.4 MHz) between the two modes implies that they are not harmonics of the vibration, rather they correspond to the non-degenerate, orthogonal flexural modes of the resonator. These orthogonal modes are shown in the inset of Figure \ref{gating}(a). The degeneracy of the orthogonal components of the fundamental mode can be broken by any asymmetry in the nanowire.
Figure \ref{gating}(b) shows the dispersion of the resonant modes with applied DC gate voltage. Mode 1 is seen to disperse negatively with increasing absolute value of the gate voltage. This is indicative of the capacitive softening \cite{kozinsky_tuning_2006,solanki_tuning_2010} effect of mode 1. Mode 2, however, is seen to have negligible dispersion with the gate voltage. The observed dispersion of the two modes to an applied DC gate voltage is indicative of their plane of oscillations. We deduce that mode 1 oscillates mainly in the plane of the substrate as the capacitive softening effect occurs in nanoelectromechanical systems (NEMS) when the oscillations are in the plane of the gate electrode. Mode 2, therefore, oscillates in a nearly vertical plane perpendicular to the substrate. Here we note that the good visibility of the modes for negative gate voltages in Figure \ref{gating}(b) arises due to the non-zero transconductance of the semiconductor nanowire. The \emph{n}-type semiconducting character of the as grown nanowires is reflected in the visibility of the dispersion diagram (see supplementary material for gating response of the nanowires).
In the following discussions we fix the gate voltage to -20 V and study the mechanical response of mode 1 to large amplitude oscillations of mode 2.

\section{Results and discussion}

\begin{figure}
\centering
\includegraphics[width=7.5cm]{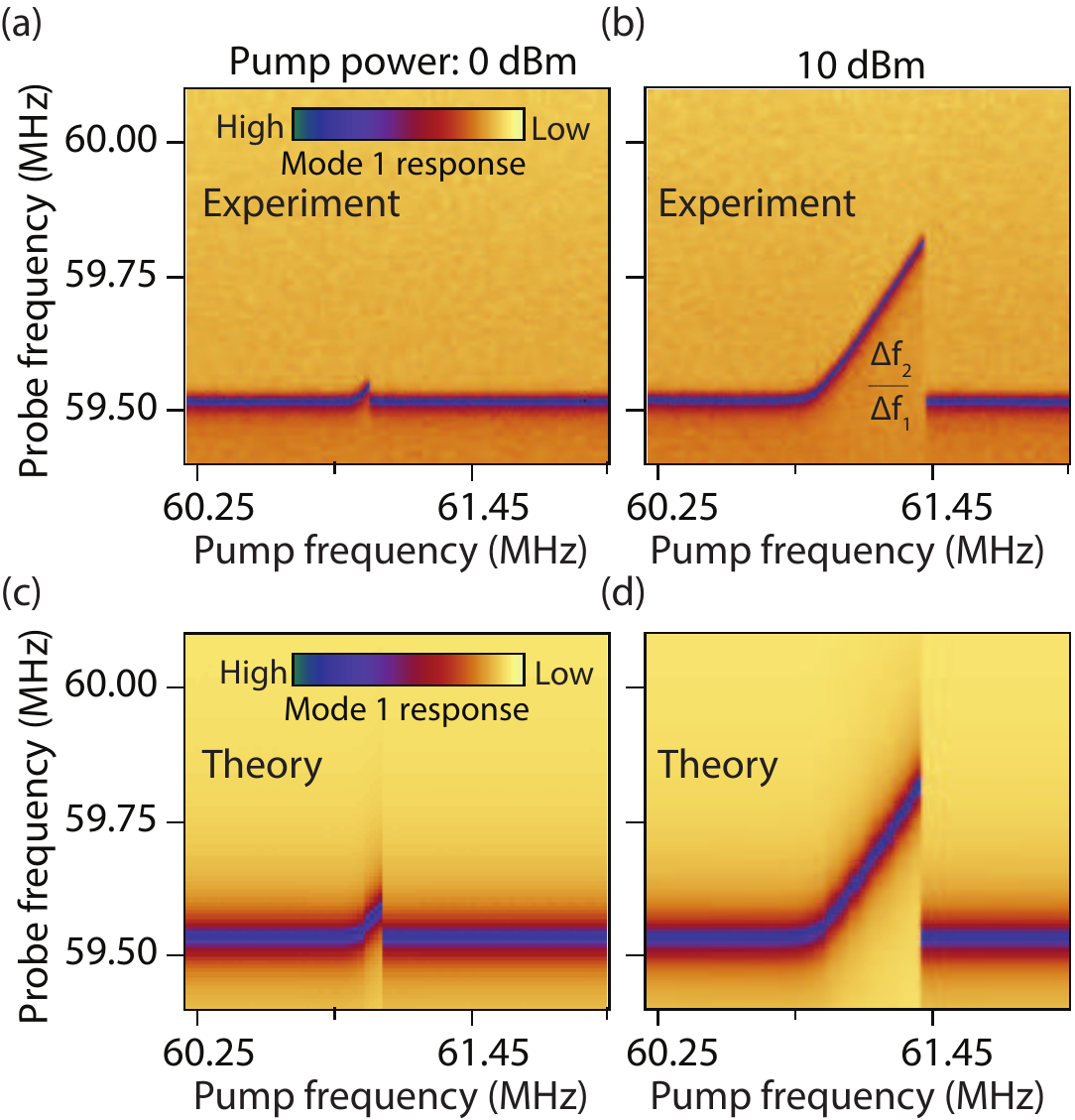}
\caption{\textbf{Nonlinear mode coupling.} \label{couplingpower} (a)-(b) Experimentally obtained response of mode 1 as a function of pump signal frequency near mode 2. The pump signal power is 0 dBm in (a) and 10 dBm in (b) where the pump frequency is increased (forward sweep) in both cases. Frequency of mode 1 increases when the pump frequency is in the range of mode 2. For larger pump power, the window of coupled response is seen to be larger. The colour scale of high/low represents mode 1 being on/off resonance. Probe power is -30 dBm. (c)-(d) Simulation showing effect of magnitude of pump signal power on frequency shift of probe response for $ \hat v_{AC1} = 0.14$ V in (c) and $ \hat v_{AC2} = 0.30$ V in (d)}
\end{figure}

Figure \ref{couplingpower} shows the effect of a large driving force on mode 2 on the frequency response of mode 1. Here the linear response of mode 1 is probed with a weak signal from the VNA with -30 dBm power whereas the pump frequency, around mode 2, is a strong signal from the FG such that mode 2 is driven in the nonlinear Duffing regime. Both the frequencies are swept in the increasing direction. For lower pump powers (0 dBm, left panel in Figure \ref{couplingpower}) where the resonant amplitude of mode 2 is estimated to be 23 nm (see supplementary material for detailed calculations), mode 1 frequency is seen to increase slightly when the pump frequency is in the vicinity of mode 2. As the pump power is increased to 10 dBm (right panel in Figure \ref{couplingpower}) mode 1 frequency is seen to be affected over larger range of the pump frequencies. As the pump frequency is increased mode 1 frequency is seen to gradually increase and abruptly decrease to its original value beyond a certain pump frequency. The observed response indicates that these orthogonal modes are nonlinearly coupled. We now explore the form of this nonlinear coupling further.

The leaning of the resonant frequency and subsequent, abrupt jump in the response is characteristic of the Duffing nonlinearity in these devices. The Duffing nonlinearity is characterized by presence of terms of the form $\alpha u^3$
in the equation of motion that gives rise to bistability in the response\cite{lifshitz2008nonlinear}. As the pump frequency matches mode 2, it drives mode 2 to large amplitudes until the point of bistability.

The response of mode 1, therefore, follows the Duffing response shape of mode 2. The frequency leaning of mode 1 arises from the deformation induced tension that accumulates on the nanowire as mode 2 oscillates with a large amplitude.
The deformation induced tension is significant when the amplitude of oscillations becomes large. This tension can be written as
\begin{equation}
\frac{EA}{2L}\int_0^L\left(\frac{\partial\epsilon}{\partial z}\right)^2dz,
\end{equation}
where $E$ is the Young's modulus, $A$ and $L$ are the cross sectional area and length of the beam respectively, and $\epsilon$ is the mode shape as a function of the distance $z$ along the length of the nanowire. The mode shape can be written in terms of the in-plane, $U(z,t)$, and out-of-plane, $V(z,t)$, components for mode 1 and mode 2 respectively. Then the governing partial differential equations of motion for each mode will include a term of the form\cite{nayfeh2008nonlinear} (see supplementary material for full Euler-Bernoulli equations)

\begin{equation}
\left[\frac{EA}{2L}\int_0^L\left(\left(\frac{\partial U(z,t)}{\partial z}\right)^2+\left(\frac{\partial V(z,t)}{\partial z}\right)^2\right)dz\right]\frac{\partial^2 W}{\partial z^2}.
\end{equation} where $W$ is either $U(z,t)$ or $V(z,t)$.

\begin{figure}
\centering
\includegraphics[width=7.5cm]{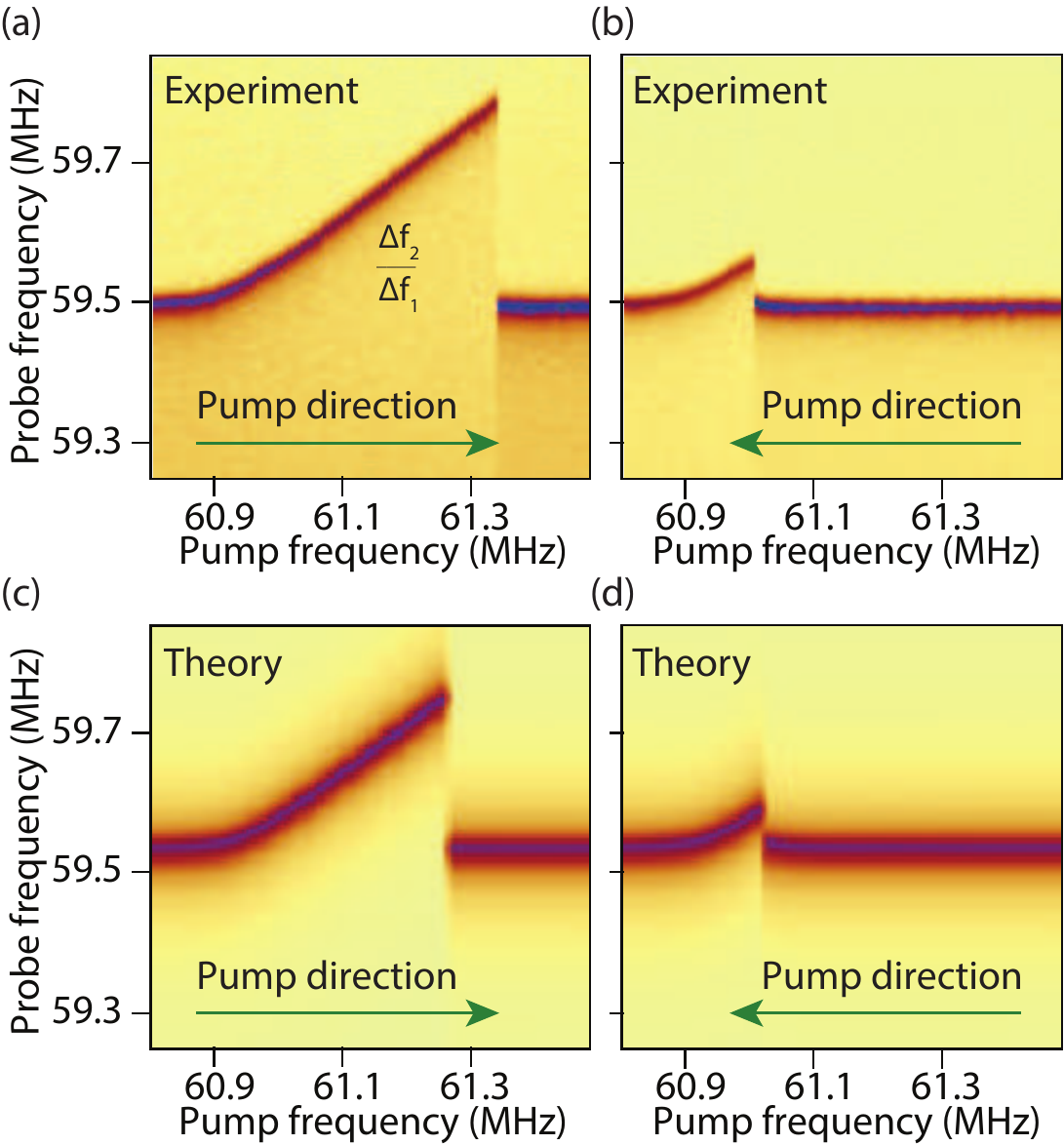}
\caption{\textbf{Hysteresis in mode coupling.} \label{couplingdirection} (a)-(b) Experimentally obtained response of mode 1 as a function of the pump signal for (a) forward and (b) reverse sweep of the pump frequency around mode 2. Probe power is -20 dBm. (c)-(d) Simulation showing hysteresis behavior in coupling regime with $ \hat v_{AC2} = 0.26$ V for forward sweep in (c) and reverse sweep in (d).}
\end{figure}

Along with the frequency leaning that accompanies the Duffing response, the bistable nature also gives rise to hysteresis in the response of a single mode with the direction of frequency sweeps. As the modes are coupled, we expect the hysteretic Duffing response of the pumped mode to reflect in the response of the probed mode as well. Figure \ref{couplingdirection} shows the effect of pump sweep direction on the response of mode 1. The coupled response of mode 1 faithfully mimics the Duffing response of mode 2.

The equations of motion for our system can be written in a simplified form using the modal coordinates, $u$ and $v$, as\cite{nayfeh2008nonlinear,eichler2012strong} (see supplementary material for details)

\begin{align}
\label{NWcouplingeqns1}
&\ddot{u}+c \dot{u}+\omega_u^2 u +\alpha_{u} u^3+\alpha_{uv} v^2 u = f_u(t)\\
&\ddot{v}+c \dot{v}+\omega_v^2 v +\alpha_{v} v^3+\alpha_{uv} u^2 v = f_v(t).
\end{align}

To arrive at equation \ref{NWcouplingeqns1} we have used a reduced order technique to solve the coupled partial differential equations (as described by Euler-Bernoulli theory) that govern the dynamics of our system. The factors $\alpha_{u}, \alpha_{v}, \alpha_{uv}\propto \alpha$ where $\alpha$ is the non-dimensional coefficient of nonlinear modal interaction governed by the geometry of the device (see supplementary material for details).
The above equations of motion capture all the physics of our system. We solve these equations numerically to compare with the experimentally observed results. The bottom panels in Figures \ref{couplingpower}-\ref{couplingdirection} show the simulated response of the nanowire resonator for corresponding experimental data in the top panels. The calculated response is seen to match well with the experimentally observed frequency shifts, bistability, and hysteresis.

In order to understand and compare our experiments with similar systems studied earlier we compare the key parameters in Table \ref{table:comparison}. $f_1$ denotes the frequency of the lower of the two coupled modes, $\Delta f \sim f_2-f_1$ is difference in the frequency of the modes and $\frac{\Delta f_2}{\Delta f_1}$ is the rate of change of frequency of the un-pumped mode as a function of the pumped mode. Larger the nonlinear coupling larger is $\frac{\Delta f_2}{\Delta f_1}$ and to first order the slope of the nonlinear coupling induced response is linear; this is seen in Figure \ref{couplingpower}(b) and Figure \ref{couplingdirection}(a). We find that the extent of nonlinear coupling is much larger in doubly clamped nanomechanical devices and this attests to the key role of tension in mediating the nonlinear coupling. Naturally, such a strong nonlinear coupling is not present in cantilevers made using nanowires \cite{cadeddu_time-resolved_2016}. We also find, from our calculations, that the coupling is the strongest between the two orthogonal modes of the fundamental (details provided in the supplementary material). It would be interesting to explore parametric degenerate and nondegenerate amplification using such coupled modes \cite{mathew_dynamical_2016}. Especially since nanowire based vectorial AFM \cite{lepinay_universal_2016,rossi_vectorial_2016} has been demonstrated using parametric schemes for these coupled modes could help push sensitivity to new limits in vectorial scanning.

\begin{table*}
\caption{Nonlinear intermodal coupling across nanomechanical systems}
\centering
\begin{tabular}{p{9.5cm}p{2cm}p{3cm}p{0.75cm}}
  \hline \hline
System used & f$_1$ & $\Delta f \sim f_2-f_1$ & $\frac{\Delta f_2}{\Delta f_1}$ \\
  \hline
Carbon nanotubes\cite{castellanos-gomez_strong_2012} (Castellanos-Gomez \emph{et al.}) & 180~MHz & $\sim$730~MHz & $\sim$1  \\
Nanofabricated doubly clamped beams \cite{westra_nonlinear_2010} (Westra \emph{et al.}) & 275~kHz & 200~kHz& $\sim$1 \\
VLS grown nanowires cantilevers \cite{cadeddu_time-resolved_2016} (Cadeddu \emph{et al.})& $\sim$1~MHz & 6~kHz & $\sim$0.1 \\
VLS grown doubly clamped nanowires (present work)  & 50~MHz & 1.6~MHz & $\sim$0.8 \\
  \hline
\end{tabular}
\label{table:comparison}
\end{table*}

\section{Conclusions}
In summary, we have studied the tension mediated nonlinear coupling between the orthogonal modes of an InAs nanowire resonator. Our experiments show that the Duffing nonlinearity of one mode affects the linear response of the second mode as observed from the abrupt and hysteretic frequency shifts. Our calculations deduce the role of large-oscillation induced tension on the observed dynamics of the system. Such coupled dynamics could be exploited to improve the performance and bandwidth of NEMS based sensors.

\section*{Acknowledgements}
We acknowledge the contribution of Mr. Mahesh Gokhale and Professor Arnab Bhattacharya, TIFR, for growth of the nanowires. We acknowledge funding from the Swarnajayanti Fellowship of DST (for MMD), Department of Atomic Energy, and Department of Science and Technology of the Government of India through the Nanomission. A.B. acknowledges the Science and Engineering Research Board, Government of India for funding.

%%%%%%%%%% Merge with supplemental materials %%%%%%%%%%
\pagebreak
\widetext
\begin{center}
\textbf{\large Supplementary material: Tension mediated nonlinear coupling between orthogonal mechanical modes of nanowire resonators}
\end{center}
%%%%%%%%%% Merge with supplemental materials %%%%%%%%%%
%%%%%%%%%% Prefix a "S" to all equations, figures, tables and reset the counter %%%%%%%%%%
\setcounter{equation}{0}
\setcounter{figure}{0}
\setcounter{table}{0}
\setcounter{section}{0}
\setcounter{page}{1}
\makeatletter
\renewcommand{\thesection}{{S}\arabic{section}}
\renewcommand{\theequation}{S\arabic{equation}}
\renewcommand{\thefigure}{S\arabic{figure}}
\renewcommand{\bibnumfmt}[1]{[S#1]}
\renewcommand{\citenumfont}[1]{S#1}

\renewcommand{\theHtable}{Supplement.\thetable}
\renewcommand{\theHfigure}{Supplement.\thefigure}
%%%%%%%%%% Prefix a "S" to all equations, figures, tables and reset the counter %%%%%%%%%%

\section{Mathematical modelling}
\begin{figure}[h]
\centering
\includegraphics[width=0.50\columnwidth]{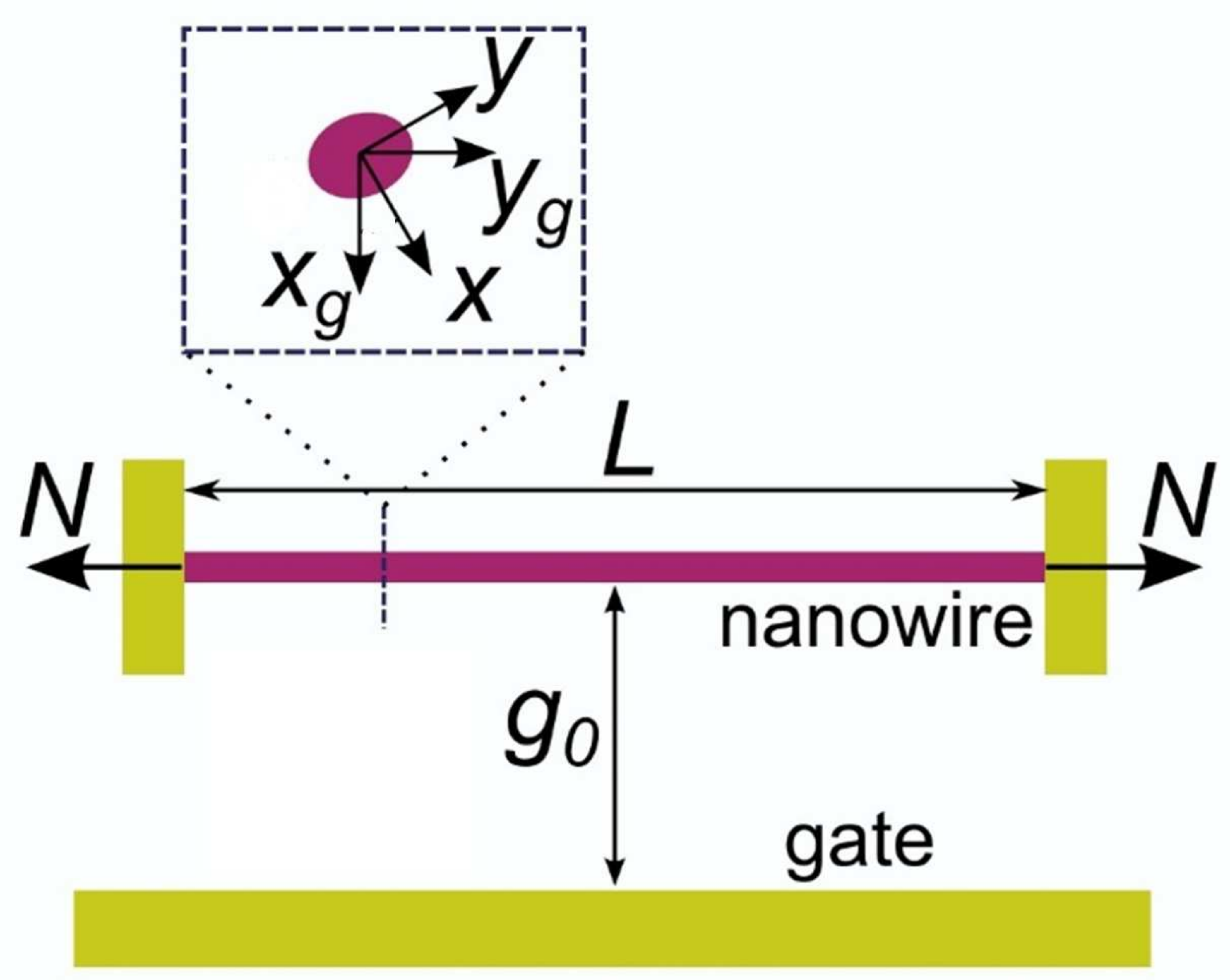}
\caption{ \label{su-nanowire}A schematic diagram of electrostatically actuated InAs nanowire.}
\end{figure}

We have performed numerical simulations to understand the mode coupling behaviour of InAs nanowire resonators. A nanowire can be modeled as cylindrical vibrating beam having same frequency in all direction of the planes of flexural vibration. But, in presence of small imperfection in cross-section, a single resonant frequency of the nanowire splits into two nearby frequencies in orthogonal principal planes \cite{santos2010}. we have observed similar two nearby resonant frequencies in the experiments of the InAs nanowire. So, the InAs nanowire has been modelled as a wire of elliptical cross-section as shown in Figure \ref{su-nanowire}. For electrostatic actuation, it is placed $g_0$ distance apart from a gate electrode. The nanowire has mass density $\rho$, length $L$, semi-major axis $r_u$, and semi-minor axis $r_v$; further, the cross-section area is denoted as $A$ and principal flexural stiffnesses as $EI_u$ and $EI_v$.
The flexural dynamics of the nanowire is described in terms of its displacements $\hat U(\hat z, \hat t)$ and $\hat V(\hat z, \hat t)$ along principal orthogonal directions (X-axis and Y-axis). Here, the variables $\hat z$ and $\hat t$ are distance coordinate along axis of the nanowire and time coordinate respectively. The purpose of using $(\hat . )$ on variables name is to distinguish their dimensional form from their non-dimensional form which are introduced in this section later. Using Euler-Bernoulli beam theory, the flexural dynamics of the electrostatically actuated InAs nanowire is governed by following coupled partial differential equations \cite{conley2008, chen2010, bhushan}
\begingroup
\small
\begin{equation}  \label{su-pde-dim}
\begin{array}{l}
 \displaystyle EI_u \frac{{\partial ^4 \hat U}}{{\partial \hat z^4 }} + \rho A\frac{{\partial ^2 \hat U}}{{\partial \hat t^2 }} + \hat c\frac{{\partial \hat U}}{{\partial \hat t}} = \left[ {\frac{{EA}}{{2L}}\int\limits_0^L {\left( {\left( {\frac{{\partial \hat U}}{{\partial \hat z}}} \right)^2  + \left( {\frac{{\partial \hat V}}{{\partial \hat z}}} \right)^2 } \right)d\hat z + \hat N} } \right]\frac{{\partial ^2 \hat U}}{{\partial \hat z^2 }} + \hat F_e^u \\
 \displaystyle EI_v \frac{{\partial ^4 \hat V}}{{\partial \hat z^4 }} + \rho A\frac{{\partial ^2 \hat V}}{{\partial \hat t^2 }} + \hat c\frac{{\partial \hat V}}{{\partial \hat t}} = \left[ {\frac{{EA}}{{2L}}\int\limits_0^L {\left( {\left( {\frac{{\partial \hat U}}{{\partial \hat z}}} \right)^2  + \left( {\frac{{\partial \hat V}}{{\partial \hat z}}} \right)^2 } \right)d\hat z + \hat N} } \right]\frac{{\partial ^2 \hat V}}{{\partial \hat z^2 }} + \hat F_e^v  \\
 \end{array}
\end{equation}
\endgroup
In Eq. \eqref{su-pde-dim}, the variable $\hat c$ is introduced to account for viscous damping, whereas the purpose of the end force $\hat N$ is to account for the presence of residual stress due to doubly-clamped or fixed boundaries. The nanowire is actuated by unit length forces $\hat F_e^u $ and $\hat F_e^v $ along x-direction and y-direction respectively. These forces build-up when we actuate the nanowire by providing DC and AC voltages between nanowire and the gate electrode. For simulations, we have assumed unit length force of an electrostatically actuated circular cross-section nanowire having radius $R=\sqrt{R_u R_v}$ as actuation force of the InAs nanowire. In case of a cylindrical nanowire, which has circular cross-section, the electrostatic actuation force is developed in the direction of gate electrode \cite{bhushan2014}
\begin{equation} \label{su-electrostat}
\hat F_e (\hat r_g, \hat t)  = \frac{{\pi \varepsilon _0 \hat V_g^2 }}{{\sqrt {\left( {g_0 + R - \hat r_g} \right)^2  - R^2 } \left[ {\cosh ^{ - 1} \left( {\frac{{g_0 + R - \hat r_g}}{R}} \right)} \right]^2 }},
\end{equation}
here $\hat r_g$ is the displacement of the nanowire in direction of the gate electrode. The variable $\hat V_g$ is the applied voltage between the nanowire and gate electrode and $\varepsilon _0$ is vacuum permittivity. In view of experiments, we have two AC voltage excitations corresponding to probe frequency $\hat \omega_{f1}$ and pump frequency $\hat \omega_{f2}$ along with the DC voltage, $\hat V_g = \left( {\hat V_{gDC}  + \hat V_{gACu} \cos (\hat \omega _{f1} \hat t) + \hat V_{gACv} \cos (\hat \omega _{f2} \hat t)} \right)$. During simulation, we have approximated $\hat F_e$ to compute unit length forces $\hat F_e^u $ along x-direction and $\hat F_e^v$ along y-direction. We have assumed the major and minor axes of the elliptical cross-section of nanowire are inclined with respect to direction of the gate electrode $x_g$ (refer Fig. \ref{su-nanowire}). The forces $\hat F_e^u = \lambda_u \hat F_e $ and $\hat F_e^v = \lambda_v \hat F_e$ are components of actuation force $\hat F_e$ of a cylindrical nanowire, where $\lambda_u$ and $\lambda_v$ are constants. \\
\indent For simplicity, we have transformed Eq. \eqref{su-pde-dim} in non-dimensional form by introducing following non-dimensional variables defined as $U = \hat U/ g_0$, $V = \hat V/ g_0$, $z = \hat z/ L$, and $t = \hat t/ T$ ($T = \sqrt{\rho A L^4/EI_u}$ is a time constant). The non-dimensional form of Eq. \ref{su-pde-dim} is
\begin{equation} \label{su-pde-nondim}
\begin{array}{l}
 U^{''''}  + \ddot U + c\dot U = \left[ {\alpha \int\limits_0^1 {\left( {U{'}^2  + V{'}^2 } \right)dz + N} } \right]U^{''} + F_e^u \\
 \gamma V^{''''}  + \ddot V + c\dot V = \left[ {\alpha \int\limits_0^1 {\left( {U{'}^2  + V{'}^2 } \right)dz + N} } \right]V^{''} + F_e^v \\
 \end{array}
\end{equation}
In Eq. \eqref{su-pde-nondim}, the notations $()^{'}$ and $\dot{()} .$ denote partial derivative with respect to $z$  and $t$ respectively. In Eq. \eqref{su-pde-nondim}, $N = \hat NL^2/EI_u$ is the non-dimensional form of the end force $\hat N$ and the variable $\gamma$ is the ratio of principal flexural stiffnesses. The non-dimensional coefficient $\alpha = Ag^2_0/2I_u$ quantifies nonlinear modal interaction behaviour in the nanowire flexural dynamics. The parameter $c$ is the non-dimensional damping coefficient. We have used the value of $c$ corresponding to experimental quality factor in numerical calculations. The non-dimensional unit length forces are denoted by $F_e^u$  and $F_e^v$. \\
\indent We have solved Eq. \eqref{su-pde-nondim} using Galerkin based reduced order model technique \cite{bhushan2014, bhushan}. The solution of Eq. \eqref{su-pde-nondim} has been assumed in form of
  \begin{equation}\label{su-assumed-sol}
U(z,t) = \phi (z)u (t)\,\,\,{\rm{and}}\,\,\,\,V(z,t) = \psi  (z)v (t).
\end{equation}
Here, $\phi(z)$ and $\psi(z)$ are mode shapes along principal directions corresponding to fundamental nearby natural frequencies, whereas $u(t)$ and $v(t)$ are modal coordinate displacements. The modeshape $\phi(z)$ (along x-direction) has been calculated by solving characteristic equation $\phi ^{''''}  = N\phi   + \omega _u^2 \phi $ and modeshape $\psi(z)$ (along y-direction) has been calculated by solving $\gamma \psi ^{''''}  = N\psi   + \omega _v^2 \psi  $. Here, $\omega_u$ and $\omega_v$ are the two nearby fundamental natural frequencies. The assumed solution \eqref{su-assumed-sol} has been substituted in Eq. \eqref{su-pde-nondim}, multiplied the equations with modeshapes, and integrated the equations from 0 to 1; we get
\begin{equation}\label{su-srom}
\begin{array}{l}
 \ddot u  + c\dot u  + \omega _u^2 u  + \alpha _u u^3  + \alpha _{uv} u v^2  = f^u_e  \\
 \ddot v  + c\dot v  + \omega _v^2 v  + \alpha _{uv} u^2 v  + \alpha _v v^3  = f^v_e \\
 \end{array}
\end{equation}
Various coefficients of the nonlinear terms of Eq. \eqref{su-srom} are
\begingroup
\footnotesize
\[
\alpha _u  = \alpha \left( {\int_0^1 {\left( {\phi{'}^2 } \right)} dz\,} \right)^2 ,\,\,\alpha _v  = \alpha \left( {\int_0^1 {\left( {\psi{'}^2 } \right)} dz\,} \right)^2 ,\,\,\alpha _{uv}  = \alpha \left( {\int_0^1 {\left( {\phi{'}^2 } \right)} dz\int_0^1 {\left( {\psi{'}^2 } \right)} dz\,} \right)
\]
\endgroup
In Eq. \eqref{su-srom}, the values of electrostatic actuation forces are $f^u_e=\int_0^1 F^e_u \phi dz$ and  $f^v_e=\int_0^1 F^e_v \psi dz$. By Taylor series expansion of the function \eqref{su-electrostat} of electrostatic forcing, one can deduce that electrostatic forcing functions $f^u_e$ and  $f^v_e$ have first harmonic excitation terms, second harmonic excitation terms, and parametric excitation terms. It is because Eq. \eqref{su-srom} is a function of displacement coordinates and due to presence of square of voltage $\hat V_g$, where $\hat V_g$ has both DC and AC voltage components. But, when investigation is near fundamental natural frequency and the gap distance between the nanowire and gate electrode is relatively larger \cite{bhushan2014, bhushan, conley2008}, first harmonic excitation term has dominant effect. So, we retain only first harmonic excitation term in electrostatic actuation force as
\begin{equation*}
f_e^u = f_u \cos(\omega_{f1} t)\,\,\, \text{and } f_e^v = f_v \cos(\omega_{f2} t)
\end{equation*}
where,
\[
\begin{array}{l}
 f_u  = 2\lambda _u f_{zero} C_{volt} \hat V_{gDC} \hat V_{gAC1} \int\limits_0^1 {\phi  } dz,\,\,\,\,\omega _{f1}  = \,\hat \omega _{f1} T, \\
 f_v  = 2\lambda _v f_{zero} C_{volt} \hat V_{gDC} \hat V_{gAC2} \int\limits_0^1 {\psi  } dz,\,\,\,\omega _{f2}  = \,\hat \omega _{f2} T, \\
\displaystyle f_{zero}  = \frac{\pi }{{\sqrt {\left( {1 + \frac{R}{{g_0 }}} \right)^2  - \left( {\frac{R}{{g_0 }}} \right)^2 } \left[ {\cosh ^{ - 1} \left( {1 + \frac{{g_0 }}{R}} \right)} \right]^2 }},\,C_{volt}  = \frac{{\varepsilon _0 L^4 }}{{g_0^2 EI_u }} \cdot \\
 \end{array}
\]

As the driving force increases, the amplitude of the modes increase giving rise to mode coupling. The amplitude of motion of the modes can be estimated by using the expression for the electrostatic driving force given in Eq. \ref{su-electrostat}. By approximating the displacement of the nanowire to be small compared to the gate electrode separation, the total amplitude of the driving force acting along the length of the nanowire can be written as 

\begin{equation*}
F=\frac{2\pi \epsilon_0 \hat{V}_{gDC}\hat{V}_{gAC} L}{\sqrt{(g_0+R)^2-R^2}\left[\cosh^{-1}\left(\frac{g_0+R}{R}\right)\right]^2}.
\end{equation*}
For a linear response, the amplitude of motion on resonance is then given by $z=\frac{F}{m} \frac{Q}{\omega_m^2}$, where $m$ is the mass of the nanowire and $\omega_m$ is the frequency of the mode. Using the dimensions of suspended length $L = 3.3$ $\mu$m, semi-minor axis $r_u = 57.7$ nm, semi-major axis $r_v = 59.3$ nm, initial gap $g_0 = 500$ nm, and quality factor $Q = 2800$, the amplitude can be estimated to be $\sim$ 0.8 nm for mode 1 for a drive power of -30 dBm, and $\sim$ 23 nm for mode 2 for a drive power of 0 dBm and DC gate voltage of 20 V.
 
\section{Numerical simulation}
\begin{figure}
\centering
\includegraphics[width=\columnwidth]{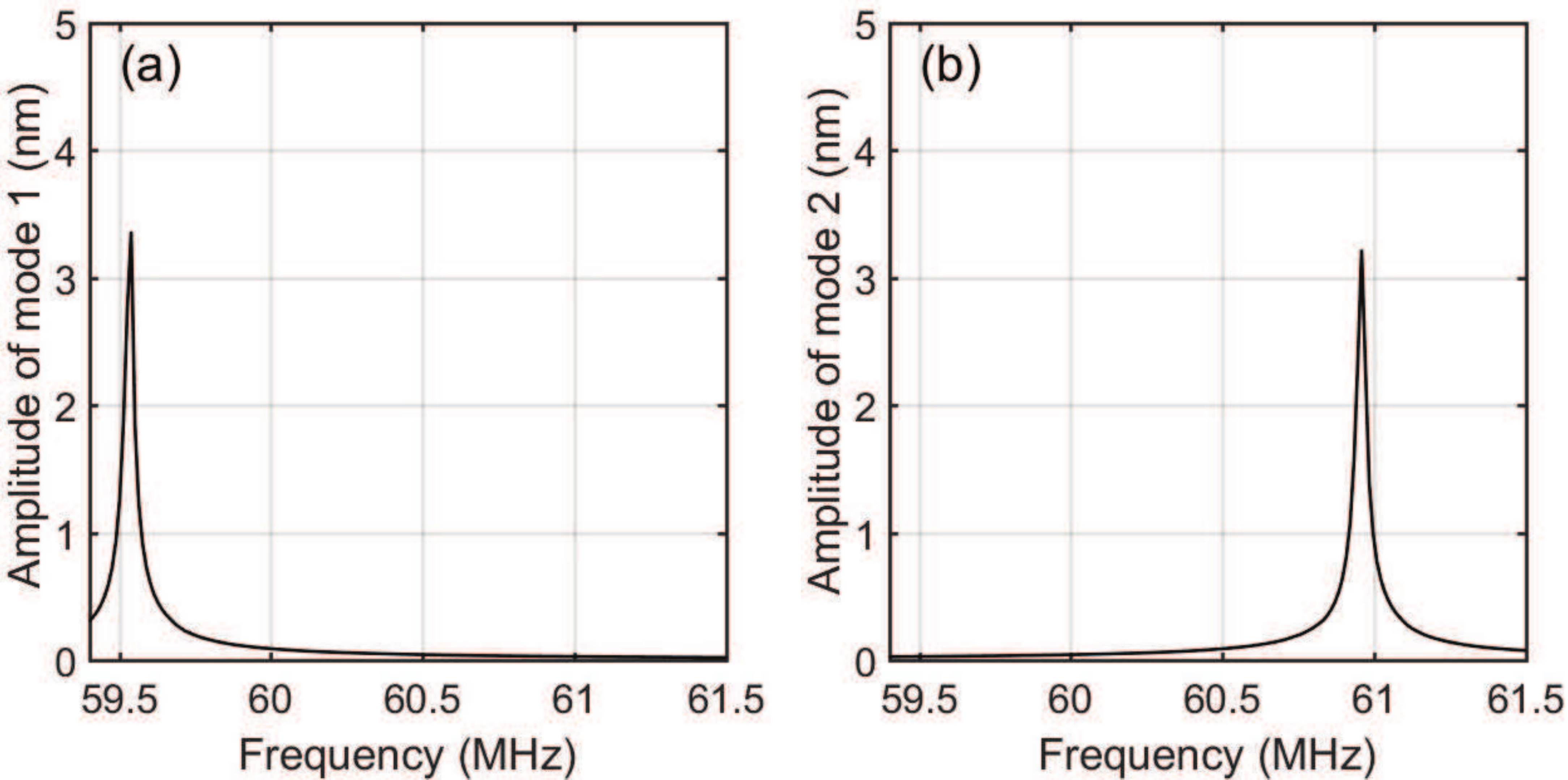}
\caption{Resonance curves near split natural frequencies (a) first mode (b) second mode.}
\label{su-resonance}
\end{figure}

\begin{figure}
\centering
\includegraphics[width=\columnwidth]{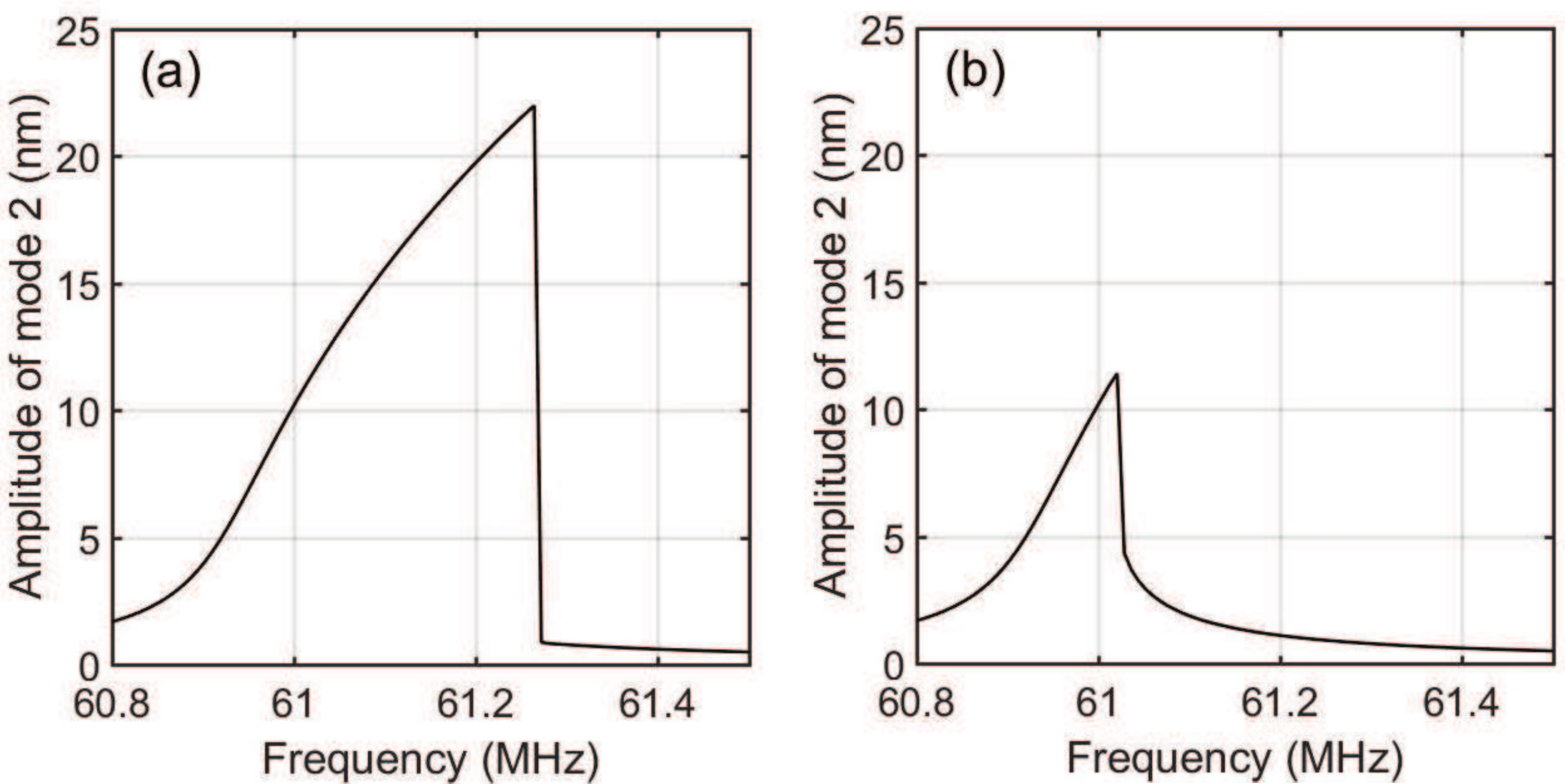}
\caption{Nonlinear resonance curve for higher magnitude of AC voltage (a) forward sweep (b) reverse sweep}
\label{su-bistable}
\end{figure}

We have solved Eq. \eqref{su-srom} to understand mode coupling behaviour of the InAs nanowire resonator. The nanowire resonator was simulated using the dimenions given above and an end force $\hat N$ as built-in tension of magnitude 1.47 times first Euler-buckling load. We have chosen the values of $r_u$ and $r_v$ near experimentally measured radius of the nanowire, along with $\hat N$, such that the nanowire has nearly same experimentally measured split resonant frequencies. In simulations, we actuate the nanowire with DC voltage $\hat V_{gDC}$ = 20 V and vary magnitude of AC voltages $\lambda_u \hat V_{gAC1} = \hat v_{ac1}$ and $\lambda_v \hat V_{gAC2} = \hat v_{ac2}$ to investigate resonance behaviour. Figures \ref{su-resonance}(a)  and \ref{su-resonance}(b) show the resonance behaviour around split resonant frequencies; here the lower and higher frequency modes are referred as mode 1 and mode 2 respectively. To obtain resonance in mode 1 oscillation, as shown in Fig. \ref{su-resonance}(a), we provide small magnitude of first AC voltage $\hat v_{ac1} = 0.04$ V in absence of second AC voltage $\hat v_{ac2} = 0$ V. Similarly, for resonance in mode 2 oscillation, as shown in Fig. \ref{su-resonance}(b), we provide second AC voltage $\hat v_{ac2} = 0.04$ V in absence of first AC voltage $\hat v_{ac1} = 0$ V. The line shape of these resonance curves demonstrate that they are linear in nature, and it is due to small magnitude of harmonic excitation. However, the nanowire resonator is inherently nonlinear in nature due to doubly-clamped boundary conditions, and nonlinearity is accounted in the equation of motion \eqref{su-srom} with Duffing nonlinearity terms. The nonlinearity effects become dominant in resonance behaviour at higher amplitude of harmonic AC voltage excitation. Figures \ref{su-bistable}(a) and \ref{su-bistable}(b) depict nonlinear resonance curves of mode 2 oscillation of the nanowire. Here, we actuate the nanowire with higher magnitude of second AC voltage $\hat v_{ac2} = 0.26$ V in absence of first AC voltage $\hat v_{ac1} = 0$ V.

The amplitude of AC voltage excitation is same for both Figs. \ref{su-bistable}(a) and \ref{su-bistable}(b), but direction of frequency sweep of AC voltage $\hat v_{ac2} \cos (\hat \omega_{f2}t)$ is different. In Fig. \ref{su-bistable}(a), the frequency sweep is in forward direction, whereas it is in reverse direction for Fig. \ref{su-bistable}(b). The resonance curves show hysteresis behaviour with change in direction of frequency sweep and it reflects the presence of nonlinearity.\\

\begin{figure}
\centering
\includegraphics[width=\columnwidth]{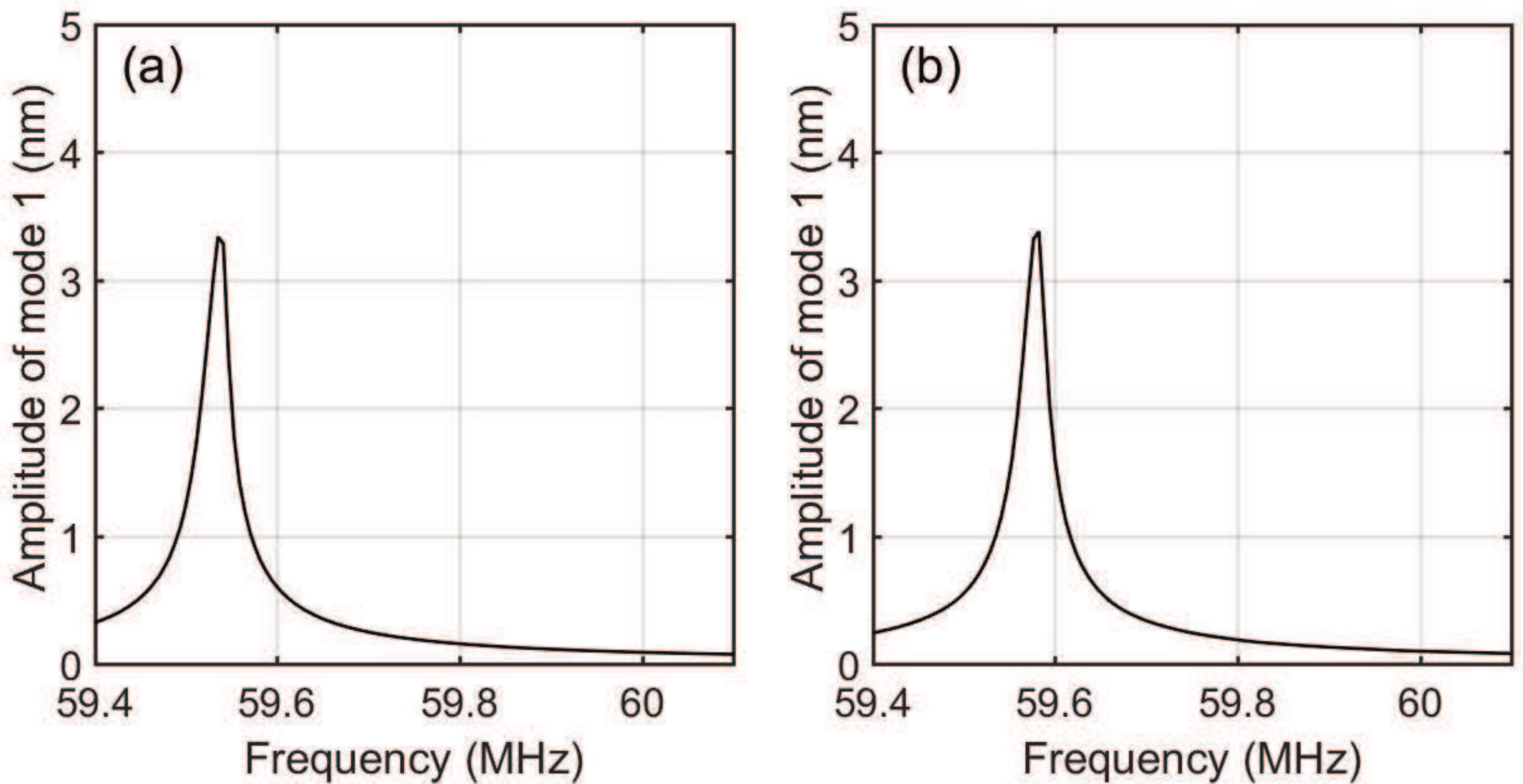}
\caption{Resonance frequency shift (a) low AC voltage excitation (b) high AC voltage excitation}
\label{su-shift}
\end{figure}

Till now, we have discussed resonance in mode 1 and mode 2 oscillations separately by providing single input AC voltage corresponding to either mode 1 frequency or mode 2 frequency in absence of other. Now, we discuss the mode coupling behaviour in the InAs nanowire by simultaneously actuating both mode 1 and mode 2 by providing both first and second AC voltages. Figures \ref{su-shift}(a) and \ref{su-shift}(b) show mode 1 resonance behaviour of the nanowire when we actuate both first and second AC voltages. Here mode 1 is probed with a small magnitude of first AC voltage $\hat v_{ac1} = 0.04$ V, and second AC voltage actuation is a pump signal. Figure \ref{su-shift}(a) shows mode 1 resonance for small magnitude of pump signal $\hat v_{ac2} = 0.04$ V. We can observe that there is shift in mode 1 resonant frequency with increase in magnitude of pump signal to $\hat v_{ac2} = 0.26$ V, as shown in Fig. \ref{su-shift}(b). In both figures, excitation frequency of pump signal is 61 MHz. We explain the frequency shift in a simplified manner in the following way. When a nanowire is placed in an undeflected position, the axial force present in the nanowire is only due to the residual stress. During oscillation, additional axial force is developed in the nanowire due to deflection under restrained boundaries, and it is measured by following nonlinear coupling factor in Eq. \eqref{su-pde-nondim}
\begin{equation} \label{su-coupling}
\alpha \int\limits_0^1 {\left( {U{'}^2  + V{'}^2 } \right)dz} \cdot
\end{equation}
The motion of the nanowire along one mode influences the motion along other mode due to this nonlinear coupling factor. In Eq. \eqref{su-pde-nondim}, first equation mainly governs mode 1 oscillation, whereas second equation governs mode 2 oscillation. When we pump mode 2 with high magnitude of AC voltage (refer second equation of Eq. \eqref{su-pde-nondim}), oscillatory motion in mode 2 vibration $V(z,t) = \psi(z)v(t)$ is set-up. The modal displacement of mode 2, $v(t)$, has harmonic response at the pump frequency $\omega_{f2} $ as  $v(t) = A_v \cos(\omega_{f2} t)$, where $A_v$ is amplitude of oscillation. Due to mode 2 oscillation, the additional axial force developed in the nanowire is proportional to square of $v(t)$ or $(A_v\cos(\omega_{f2}t))^2$ or $(A_v^2/2)(1 + \cos(2 \omega_{f2} t))$. The additional axial force affects the mode 1 resonance due to presence of the nonlinear coupling factor \eqref{su-coupling}.  Hence, the constant component of additional axial force, proportional to $A_v^2/2$, adds up with the residual axial load and eventually shifts the resonant frequency of mode 1 vibration (refer first equation of Eq. \eqref{su-srom}).
\section{Analytical solution}
\begin{figure}
\centering
\includegraphics[width=\columnwidth]{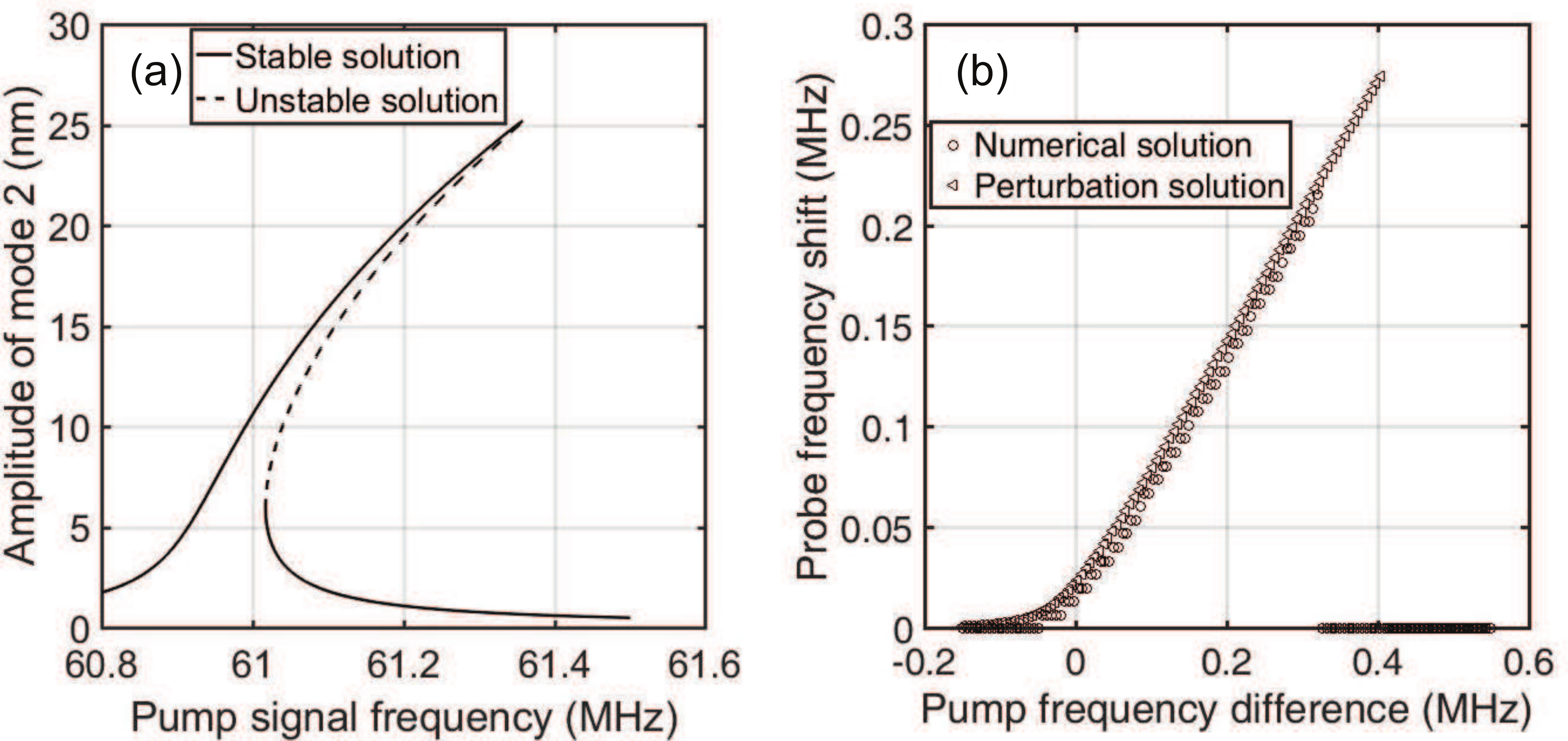}
\caption{(a) Nonlinear resonance curve of mode 2 vibration when magnitude of second AC voltage $\hat v_{ac2} = 0.26$ V (b) Relationship between frequency tuning in probe signal $\Delta_{probe}$ with frequency difference of pump signal $\Delta_{pump}$ due to nonlinear coupling.}
\label{su-Nresonance}
\end{figure}

We provide analytical expressions for computing the resonant frequency shift in probe mode oscillation. The magnitude of this resonant frequency shift can be computed by expression
\begin{equation}\label{su-frequency-shift}
\Delta_{probe}  = \left( {\sqrt {\omega _u^2  + \alpha _{uv} \frac{{A_v^2 }}{2}} } \right) - \omega _u.
\end{equation}
Further, the amplitude of mode 2 oscillation $A_v$ can be obtained analytically by solving equation of motion using perturbation method under following approximations. In the present investigation, mode 1 is excited with a low magnitude AC voltage in comparison to mode 2 excitation. So, when we solve Eq. \eqref{su-srom}, the effect of mode 1 vibration is negligible in computation of amplitude of mode 2, $A_v$. We have solved the second equation of Eq. \eqref{su-srom}, after ignoring coupling terms $\alpha _{uv} u^2 v$, using the perturbation technique, method of multiple scales \cite{nayfeh2008nonlinear}. The perturbation solution can be presented as a frequency response equation to compute the amplitude $A_v$ as a function of frequency difference in pump signal $\Delta_{pump} = \omega_{f2} - \omega_{v}$, and it is given by
\begin{equation}\label{su-perturbation}
\left( {\frac{c}{2}} \right)^2  + \left( {\Delta_{pump}  - \frac{{3\alpha _v A_v^2 }}{{8\omega _v }}} \right)^2  = \frac{{f_v^2 }}{{4\omega _v^2 A_v^2 }}.
\end{equation}
Hence, by solving Eqs. \eqref{su-frequency-shift} and \eqref{su-perturbation}, we can obtain probe signal frequency shift $\Delta_{probe}$ as a function of pump signal frequency difference $\Delta_{pump}$.\\
\indent We present the effectiveness of Eqs. \eqref{su-frequency-shift} and \eqref{su-perturbation} in computation of frequency shift $\Delta_{probe}$ using Figs. \ref{su-Nresonance}(a) and (b). Here, Fig. \ref{su-Nresonance}(a) shows nonlinear resonance behaviour of mode 2 oscillation when magnitude of pump signal or second AC voltage $\hat v_{ac2} = 0.26$ V; it is obtained by solving Eq. \eqref{su-perturbation}. By comparing Fig. \ref{su-Nresonance}(a) with Figs. \ref{su-bistable}(a) and (b), we can say that perturbation solution \eqref{su-perturbation} is in good agreement with numerical solution of Eq. \eqref{su-srom}. We have further computed frequency shift $\Delta_{probe}$ using solution of Eq. \eqref{su-perturbation} and Eq. \eqref{su-frequency-shift}, and the solution is presented here in Fig. \ref{su-Nresonance}(b) by showing relationship between $\Delta_{probe}$ and $\Delta_{pump}$. In this figure, we have also compared the analytical solution of Eq. \eqref{su-frequency-shift} with the numerical solution of Eq. \eqref{su-srom}. There is good agreement between both solutions, and it demonstrates that the analytical expressions can serve as useful tool for computing frequency shift $\Delta_{probe}$.\\
\indent In this work, modal interaction between first planar and first nonplanar modes of vibration has been investigated. However, researchers have earlier investigated similar modal interaction between first planar and higher planar modes \cite{westra2010, gomez2012, westra2011}. The derived expressions Eqs. \eqref{su-frequency-shift} and \eqref{su-perturbation} are also useful for obtaining the relationship between probe mode frequency shift $\Delta_{probe}$ (first planar mode) and pump mode frequency difference  $\Delta_{pump}$ (higher planar mode). The motion of such a system is only governed by the first equation of Eq. \eqref{su-pde-nondim} because modal interaction between only planar modes are under consideration. From Eq. \eqref{su-pde-nondim}, we get a system of two second order ordinary differential equations like Eq. \eqref{su-srom} by following Galerkin procedure as applied for derivation of Eq. \eqref{su-srom}. Further, Eqs. \eqref{su-frequency-shift} and \eqref{su-perturbation} are applicable to obtain relationship between $\Delta_{probe}$ and $\Delta_{pump}$. 

\indent To compare the modal interaction between different modes using Eqs. \eqref{su-frequency-shift} and \eqref{su-perturbation} we define a tuning constant $K_{tuning} = \alpha_{uv}f_v^2/c^2\omega_v^2$ which approximately quantifies relative coupling strength in modal interaction. The motivation in defining this constant is as follow. We can expand expression Eq. \eqref{su-frequency-shift} using Taylor series expansion and retain only linear term, and get that $\Delta_{probe}$ is directly proportional to $\alpha_{uv}$ and $A_{v}^2$. In linear resonance case, the amplitude at resonance $A_v$ is directly proportional to $f_v$ and inversely proportional to $c$ and $\omega_v$. Next, we have computed tuning constant $K_{tuning}$ for different cases of modal interaction in Table \ref{su-table}. Here, we consider three cases of modal interaction for first planar mode with first nonplanar mode (Type 1), first planar mode with second planar mode (Type 2), and first planar mode with third planar mode (Type 3). It is interesting that tuning constant is highest for the case of Type 1 modal interaction. The tuning constant for Type 2 modal interaction is zero due to antisymmetric nature of second mode which makes $f_v=0$. We can also deduce from Table \ref{su-table} that tuning constant $K_{tuning}$ for Type 1 modal interaction is about 2.65 times greater than Type 3 modal interaction. 

\begin{table}
\begin{tabular}{|c|c|c|c|}
  \hline
   & First planar and  & First planar and  & First planar and  \\
   &  first nonplanar & second planar & third planar \\
   &  (Type 1) & (Type 2) & (Type 3) \\
  \hline
  Tuning constant & 22.13 & 0 & 8.33 \\
  ($K_{tuning}$) & & &\\
  \hline
\end{tabular}
\caption{Magnitude of tuning constant $K_{tuning}$ for different types of modal interaction}
\label{su-table}
\end{table}

\begin{figure}
\centering
\includegraphics[width=\columnwidth]{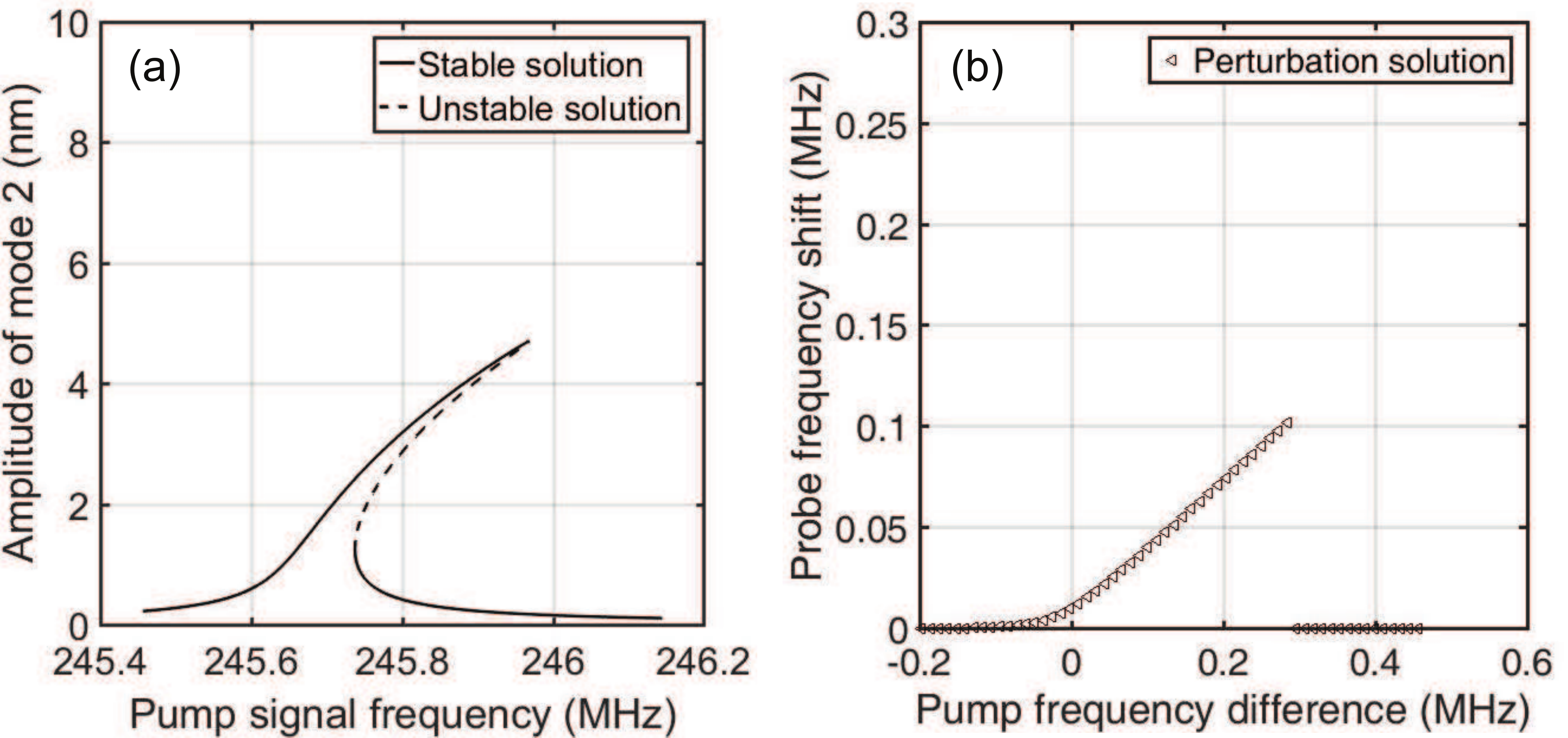}
\caption{(a) Nonlinear resonance curve of planar third mode vibration when magnitude of second AC voltage $\hat v_{ac2} = 0.26$ V (b) Relationship between frequency tuning in probe signal $\Delta_{probe}$ (planar first mode) with frequency difference of pump signal $\Delta_{pump}$ (planar third mode) due to nonlinear coupling.}
\label{su-NresonanceHigher}
\end{figure}

\indent Relationship between $\Delta_{probe}$ and $\Delta_{pump}$ for case of Type 1 modal interaction is already shown in Fig. \ref{su-Nresonance}(b). We have further solved Eqs.\eqref{su-frequency-shift} and \eqref{su-perturbation} for investigating interaction of Type 3 modal interaction. We present the solution here in Figs. \ref{su-NresonanceHigher}(a) and (b). Here, Fig. \ref{su-NresonanceHigher}(a) shows nonlinear resonance behaviour of third planar mode oscillation when magnitude of pump signal or second AC voltage $\hat v_{ac2} = 0.26$ V as in case for Fig. \ref{su-Nresonance}(a). Further, the relationship between $\Delta_{probe}$ and $\Delta_{pump}$ for Type 3 modal interaction is shown in Fig. \ref{su-NresonanceHigher}(b). Under similar loading condition, we can compare coupling strength of Type 1 and Type 3 modal interaction by observing Figs. \ref{su-Nresonance}(b) and \ref{su-NresonanceHigher}(b). From these figures, we can deduce that tuning capacity (maximum value of $\Delta_{probe}$) in Type 1 modal interaction is about 2.65 times than Type 3 modal interaction. It is consistent with comparison of tuning constant $K_{tuning}$ of Type 1 and Type 3 modal interactions.

\begin{figure}[h]
\centering
\includegraphics[width=0.5\columnwidth]{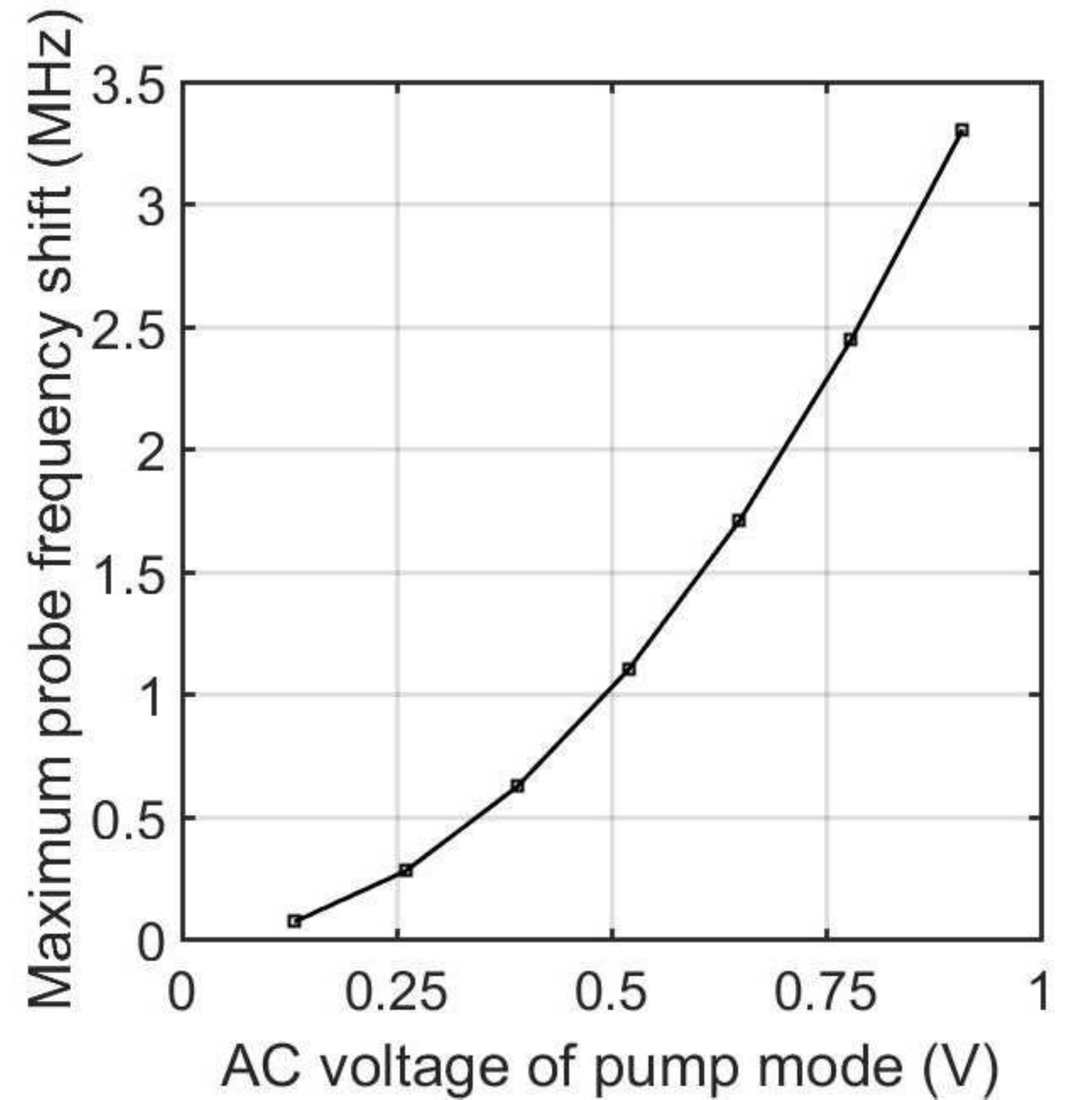}
\caption{ \label{su-VAC2coupling} Effect of pump mode AC voltage on the maximum frequency shift of the probe mode.}
\end{figure}
\indent We can ascertain more information on Type 1 modal interaction by analyzing the expression for the tuning constant, $K_{tuning}= \alpha_{uv}f_v^2/c^2\omega_v^2$. It provides insight on the frequency shift behaviour of the probe mode $\Delta_{probe}$ for varying magnitude of AC voltage of the pump mode $\hat v_{ac2}$. The maximum value of frequency shift is approximately proportional to the square of the magnitude of $\hat v_{ac2}$. This is because $K_{tuning}$ is proportional to the square of $f_v$, and $f_v$ is proportional to $\hat v_{ac2}$. Figure \ref{su-VAC2coupling} shows the calculated dependence of maximum frequency shift of probe mode on the magnitude of AC voltage of the pump mode.

\section{Gating response}
\begin{figure}
\centering
\includegraphics[width=0.67\columnwidth]{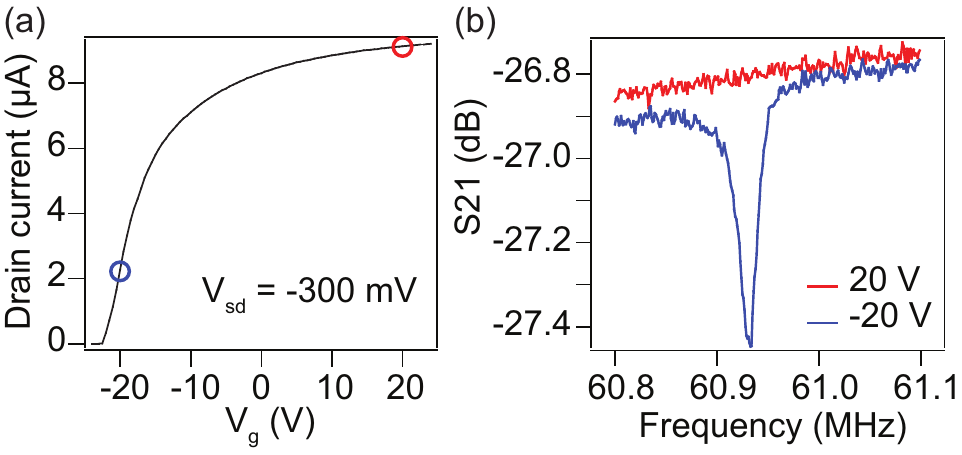}
\caption[Gating response and resonance detection]{\label{transconductance} (a) The n-type field effect transistor response of the semiconducting nanowire is seen from the variation of the DC current through the nanowire as a function of the DC gate voltage. The slope of the curve is directly proportional to the transconductance of the device. The blue circle marks a region of high transconductance whereas the red circle marks a region of low transconductance. (b) The response of mode 2 measured at the gate voltages marked by the circles in (a) shows the effect of transconductance on the measured \textit{rf} signal.}
\end{figure}

The semiconducting nanowires used in this work are \emph{n}-type in nature. As the conductance of the device depends on the carrier density in the nanowire the gate voltage can deplete or saturate the conductance (see Figure \ref{transconductance}(a)). Therefore, as the nanowire undergoes mechanical oscillations the conductance of the device itself is modified due to the modulated gate capacitance. The conductance modulation due to oscillations is large when the transconductance of the device is large. Therefore, the strength of measured resonance signal depends on the transconductance.

Figure \ref{transconductance} shows the resonant response of mode 2 of the nanowire resonator at $\pm$ 20 V gate voltage showing the effect of a non-zero transconductance. As the magnitude of the gate voltage is same, the force acting on the resonator in both cases are equal. However, in regions (blue circle) where the transconductance of the device is high our measurement scheme is more effective in detecting the mechanical motion. Typical resonant response measured in experiments is seen in Figure \ref{transconductance}(b) as a dip. 

\begin{thebibliography}{10}
\expandafter\ifx\csname url\endcsname\relax
  \def\url#1{\texttt{#1}}\fi
\expandafter\ifx\csname urlprefix\endcsname\relax\def\urlprefix{URL }\fi
\expandafter\ifx\csname href\endcsname\relax
  \def\href#1#2{#2} \def\path#1{#1}\fi

\bibitem{aspelmeyer_cavity_2014}
M.~Aspelmeyer, T.~J. Kippenberg, F.~Marquardt,
  \href{http://link.aps.org/doi/10.1103/RevModPhys.86.1391}{Cavity
  optomechanics}, Reviews of Modern Physics 86~(4) (2014) 1391--1452.
%\newblock \href {http://dx.doi.org/10.1103/RevModPhys.86.1391}



\bibitem{mahboob_phonon-cavity_2012}
I.~Mahboob, K.~Nishiguchi, H.~Okamoto, H.~Yamaguchi,
  \href{http://www.nature.com/nphys/journal/v8/n5/full/nphys2277.html}{Phonon-cavity
  electromechanics}, Nature Physics 8~(5) (2012) 387--392.
%\newblock \href {http://dx.doi.org/10.1038/nphys2277}



\bibitem{okamoto_coherent_2013}
H.~Okamoto, A.~Gourgout, C.-Y. Chang, K.~Onomitsu, I.~Mahboob, E.~Y. Chang,
  H.~Yamaguchi,
  \href{http://www.nature.com/nphys/journal/v9/n8/full/nphys2665.html}{Coherent
  phonon manipulation in coupled mechanical resonators}, Nature Physics 9~(8)
  (2013) 480--484.
%\newblock \href {http://dx.doi.org/10.1038/nphys2665}



\bibitem{faust_nonadiabatic_2012}
T.~Faust, J.~Rieger, M.~J. Seitner, P.~Krenn, J.~P. Kotthaus, E.~M. Weig,
  \href{https://link.aps.org/doi/10.1103/PhysRevLett.109.037205}{Nonadiabatic
  {Dynamics} of {Two} {Strongly} {Coupled} {Nanomechanical} {Resonator}
  {Modes}}, Physical Review Letters 109~(3) (2012) 037205.
%\newblock \href {http://dx.doi.org/10.1103/PhysRevLett.109.037205}



\bibitem{westra_interactions_2011}
H.~J.~R. Westra, D.~M. Karabacak, S.~H. Brongersma, M.~Crego-Calama, H.~S.~J.
  van~der Zant, W.~J. Venstra,
  \href{https://link.aps.org/doi/10.1103/PhysRevB.84.134305}{Interactions
  between directly- and parametrically-driven vibration modes in a
  micromechanical resonator}, Physical Review B 84~(13) (2011) 134305.
%\newblock \href {http://dx.doi.org/10.1103/PhysRevB.84.134305}



\bibitem{westra_nonlinear_2010}
H.~J.~R. Westra, M.~Poot, H.~S.~J. van~der Zant, W.~J. Venstra,
  \href{http://link.aps.org/doi/10.1103/PhysRevLett.105.117205}{Nonlinear
  {Modal} {Interactions} in {Clamped}-{Clamped} {Mechanical} {Resonators}},
  Physical Review Letters 105~(11) (2010) 117205.
%\newblock \href {http://dx.doi.org/10.1103/PhysRevLett.105.117205}



\bibitem{castellanos-gomez_strong_2012}
A.~Castellanos-Gomez, H.~B. Meerwaldt, W.~J. Venstra, H.~S.~J. van~der Zant,
  G.~A. Steele,
  \href{https://link.aps.org/doi/10.1103/PhysRevB.86.041402}{Strong and tunable
  mode coupling in carbon nanotube resonators}, Physical Review B 86~(4) (2012)
  041402.
%\newblock \href {http://dx.doi.org/10.1103/PhysRevB.86.041402}



\bibitem{eichler2012strong}
A.~Eichler, M.~del {\'A}lamo~Ruiz, J.~Plaza, A.~Bachtold, Strong coupling
  between mechanical modes in a nanotube resonator, Physical {Review Letters}
  109~(2) (2012) 025503.

\bibitem{alba_tunable_2016}
R.~D. Alba, F.~Massel, I.~R. Storch, T.~S. Abhilash, A.~Hui, P.~L. McEuen,
  H.~G. Craighead, J.~M. Parpia,
  \href{https://www.nature.com/articles/nnano.2016.86}{Tunable phonon-cavity
  coupling in graphene membranes}, Nature Nanotechnology 11~(9) (2016)
  741--746.
%\newblock \href {http://dx.doi.org/10.1038/nnano.2016.86}



\bibitem{mathew_dynamical_2016}
J.~P. Mathew, R.~N. Patel, A.~Borah, R.~Vijay, M.~M. Deshmukh,
  \href{https://www.nature.com/articles/nnano.2016.94}{Dynamical strong
  coupling and parametric amplification of mechanical modes of graphene drums},
  Nature Nanotechnology 11~(9) (2016) 747--751.
%\newblock \href {http://dx.doi.org/10.1038/nnano.2016.94}



\bibitem{liu_optical_2015}
C.-H. Liu, I.~S. Kim, L.~J. Lauhon,
  \href{http://dx.doi.org/10.1021/acs.nanolett.5b02586}{Optical {Control} of
  {Mechanical} {Mode}-{Coupling} within a {MoS}$_2$ {Resonator} in the
  {Strong}-{Coupling} {Regime}}, Nano Letters 15~(10) (2015) 6727--6731.
%\newblock \href {http://dx.doi.org/10.1021/acs.nanolett.5b02586}



\bibitem{cadeddu_time-resolved_2016}
D.~Cadeddu, F.~R. Braakman, G.~T\"ut\"unc\"uoglu, F.~Matteini, D.~R\"uffer,
  A.~Fontcuberta~i Morral, M.~Poggio,
  \href{http://dx.doi.org/10.1021/acs.nanolett.5b03822}{Time-{Resolved}
  {Nonlinear} {Coupling} between {Orthogonal} {Flexural} {Modes} of a
  {Pristine} {GaAs} {Nanowire}}, Nano Letters 16~(2) (2016) 926--931.
%\newblock \href {http://dx.doi.org/10.1021/acs.nanolett.5b03822}



\bibitem{lepinay_universal_2016}
L.~M.~d. L\'effpinay, B.~Pigeau, B.~Besga, P.~Vincent, P.~Poncharal,
  O.~Arcizet, \href{https://www.nature.com/articles/nnano.2016.193}{A universal
  and ultrasensitive vectorial nanomechanical sensor for imaging 2d force
  fields}, Nature Nanotechnology 12~(2) (2016) 156--162.
%\newblock \href {http://dx.doi.org/10.1038/nnano.2016.193}



\bibitem{foster_tuning_2016}
A.~P. Foster, J.~K. Maguire, J.~P. Bradley, T.~P. Lyons, A.~B. Krysa, A.~M.
  Fox, M.~S. Skolnick, L.~R. Wilson,
  \href{http://dx.doi.org/10.1021/acs.nanolett.6b02994}{Tuning {Nonlinear}
  {Mechanical} {Mode} {Coupling} in {GaAs} {Nanowires} {Using}
  {Cross}-{Section} {Morphology} {Control}}, Nano Letters 16~(12) (2016)
  7414--7420.
%\newblock \href {http://dx.doi.org/10.1021/acs.nanolett.6b02994}



\bibitem{rossi_vectorial_2016}
N.~Rossi, F.~R. Braakman, D.~Cadeddu, D.~Vasyukov, G.~T\"ut\"unc\"uoglu, A.~F.~i.
  Morral, M.~Poggio,
  \href{https://www.nature.com/articles/nnano.2016.189}{Vectorial scanning
  force microscopy using a nanowire sensor}, Nature Nanotechnology 12~(2)
  (2016) 150--155.
%\newblock \href {http://dx.doi.org/10.1038/nnano.2016.189}



\bibitem{solanki_tuning_2010}
H.~S. Solanki, S.~Sengupta, S.~Dhara, V.~Singh, S.~Patil, R.~Dhall, J.~Parpia,
  A.~Bhattacharya, M.~M. Deshmukh,
  \href{http://link.aps.org/doi/10.1103/PhysRevB.81.115459}{Tuning mechanical
  modes and influence of charge screening in nanowire resonators}, Physical
  Review B 81~(11) (2010) 115459.
%\newblock \href {http://dx.doi.org/10.1103/PhysRevB.81.115459}



\bibitem{lifshitz2008nonlinear}
R.~Lifshitz, M.~Cross, Nonlinear dynamics of nanomechanical and micromechanical
  resonators, Review of nonlinear dynamics and complexity 1 (2008) 1--52.

\bibitem{khan_tension-induced_2013}
R.~Khan, F.~Massel, T.~T. Heikkil\"a,
  \href{http://link.aps.org/doi/10.1103/PhysRevB.87.235406}{Tension-induced
  nonlinearities of flexural modes in nanomechanical resonators}, Physical
  Review B 87~(23) (2013) 235406.
%\newblock \href {http://dx.doi.org/10.1103/PhysRevB.87.235406}



\bibitem{eriksson_frequency_2013}
A.~M. Eriksson, D.~Midtvedt, A.~Croy, A.~Isacsson,
  \href{http://stacks.iop.org/0957-4484/24/i=39/a=395702}{Frequency tuning,
  nonlinearities and mode coupling in circular mechanical graphene resonators},
  Nanotechnology 24~(39) (2013) 395702.
%\newblock \href {http://dx.doi.org/10.1088/0957-4484/24/39/395702}



\bibitem{dasgupta_25th_2014}
N.~P. Dasgupta, J.~Sun, C.~Liu, S.~Brittman, S.~C. Andrews, J.~Lim, H.~Gao,
  R.~Yan, P.~Yang,
  \href{http://onlinelibrary.wiley.com/doi/10.1002/adma.201305929/abstract}{25th
  {Anniversary} {Article}: {Semiconductor} {Nanowires} - {Synthesis},
  {Characterization}, and {Applications}}, Advanced Materials 26~(14) (2014)
  2137--2184.
%\newblock \href {http://dx.doi.org/10.1002/adma.201305929}



\bibitem{algra_twinning_2008}
R.~E. Algra, M.~A. Verheijen, M.~T. Borgstr\"om, L.-F. Feiner, G.~Immink,
  W.~J.~P. van Enckevort, E.~Vlieg, E.~P. A.~M. Bakkers,
  \href{http://dx.doi.org/10.1038/nature07570}{Twinning superlattices in indium
  phosphide nanowires}, Nature 456 (2008) 369.


\bibitem{johansson_controlled_2009}
P.~Caroff, K.~A. Dick, J.~Johansson, M.~E. Messing, K.~Deppert, L.~Samuelson,
  Controlled polytypic and twin-plane superlattices in iii--v nanowires, Nature
  Nanotechnology 4~(1) (2009) 50--55.

\bibitem{abhilash2012}
T.~Abhilash, J.~P. Mathew, S.~Sengupta, M.~R. Gokhale, A.~Bhattacharya, M.~M.
  Deshmukh, Wide bandwidth nanowire electromechanics on insulating substrates
  at room temperature, {Nano Letters} 12~(12) (2012) 6432--6435.

\bibitem{xu2010radio}
Y.~Xu, C.~Chen, V.~V. Deshpande, F.~A. DiRenno, A.~Gondarenko, D.~B. Heinz,
  S.~Liu, P.~Kim, J.~Hone, Radio frequency electrical transduction of graphene
  mechanical resonators, Applied Physics Letters 97~(24) (2010) 243111.

\bibitem{kozinsky_tuning_2006}
I.~Kozinsky, H.~W.~C. Postma, I.~Bargatin, M.~L. Roukes,
  \href{http://scitation.aip.org/content/aip/journal/apl/88/25/10.1063/1.2209211}{Tuning
  nonlinearity, dynamic range, and frequency of nanomechanical resonators},
  Applied Physics Letters 88~(25) (2006) 253101.
%\newblock \href {http://dx.doi.org/10.1063/1.2209211}



\bibitem{nayfeh2008nonlinear}
A.~H. Nayfeh, D.~T. Mook, Nonlinear oscillations, John Wiley \& Sons, 2008.

\end{thebibliography}

\clearpage

\begin{thebibliography}{1}
\expandafter\ifx\csname url\endcsname\relax
  \def\url#1{\texttt{#1}}\fi
\expandafter\ifx\csname urlprefix\endcsname\relax\def\urlprefix{URL }\fi
\expandafter\ifx\csname href\endcsname\relax
  \def\href#1#2{#2} \def\path#1{#1}\fi

\bibitem{santos2010}
E.~Gil-santos, D.~Ramos, J.~Martinez, M.~Fernandez-Regulez, R.~Garcia, A.~S.
  Paulo, M.~Calleja, J.~Tamayo, Nanomechanical mass sensing and stiffness
  spectrometry based on two-dimensional vibrations of resonant nanowires,
  Nature Nanotechnology 5 (2010) 641--645.

\bibitem{conley2008}
W.~G. Conley, A.~Raman, C.~M. Krousgrill, S.~Mohammadi, Nonlinear and nonplanar
  dynamics of suspended nanotube and nanowire resonators, Nano Letters 8 (2008)
  1590--1595.

\bibitem{chen2010}
Q.~Chen, L.~Huang, Y.~Lai, C.~Grebogi, D.~Dietz, Extensively chaotic motion in
  electrostatically driven nanowires and applications, Nano Letters 10 (2010)
  406--413.

\bibitem{bhushan}
A.~Bhushan, M.~M. Inamdar, D.~N. Pawaskar, Effects of {DC} voltage on
  initiation of whirling motion of an electrostatically actuated nanowire
  oscillator, arXiv:1307.2359 [cond-mat.mes-hall].

\bibitem{bhushan2014}
A.~Bhushan, M.~M. Inamdar, D.~N. Pawaskar, Simultaneous planar free and forced
  vibrations analysis of an electrostatically actuated beam oscillator,
  International Journal of Mechanical Sciences 82 (2014) 90--99.

\bibitem{nayfeh2008nonlinear}
A.~H. Nayfeh, D.~T. Mook, Nonlinear oscillations, John Wiley \& Sons, 2008.

\bibitem{westra2010}
H.~J.~R. Westra, M.~Poot, H.~S.~J. van~der Zant, W.~J. Venstra, Nonlinear modal
  interactions in clamped-clamped mechanical resonators, Physical Review
  Letters 105 (2010) 117205.

\bibitem{gomez2012}
A.~Castellanos-Gomez, H.~B. Meerwaldt, W.~J. Venstra, H.~S.~J. van~der Zant,
  G.~A. Steele, Strong and tunable mode coupling in carbon nanotube resonators,
  Physical Review {B} 86 (2012) 041402.

\bibitem{westra2011}
H.~J.~R. Westra, D.~M. Karabacak, S.~H. Brongersma, M.~Crego-Calama, H.~S.~J.
  van~der Zant, W.~J. Venstra, Interactions between directly- and
  parametrically-driven vibration modes in a micromechanical resonator,
  Physical Review {B} 84 (2011) 134305.

\end{thebibliography}
\end{document}